\documentclass[preprint,noabstract,preprintnumbers,amsmath,amssymb]{revtex4}
\usepackage{graphicx}
\usepackage{dcolumn}
\usepackage{bm}

\begin{document}
\title{Coexistence of Magnetic Order and Two-dimensional Superconductivity at LaAlO$_3$/SrTiO$_3$ Interfaces}
\author{Lu Li$^1$, C. Richter$^2$, J. Mannhart$^2$, R. C. Ashoori$^1$
}
\affiliation{
$^1$ Department of Physics, Massachusetts Institute of Technology, Cambridge, Massachusetts 02139, USA.\\
$^2$ Center for Electronic Correlations and Magnetism, University of Augsburg, Augsburg, 86135, Germany.
}
\date{\today}



\maketitle                   

{\bf A two dimensional electronic system with novel electronic properties forms at the interface between the insulators LaAlO$_3$ and SrTiO$_3$~\cite{HwangNature2004,MannhartScience2006}. Samples fabricated until now have been found to be either magnetic or superconducting, depending on growth conditions. We combine transport measurements with high-resolution magnetic torque magnetometry and report here evidence of magnetic ordering of the two-dimensional electron liquid at the interface. The magnetic ordering exists from well below the superconducting transition to up to 200 K, and is characterized by an in-plane magnetic moment. Our results suggest that there is either phase separation or coexistence between magnetic and superconducting states. The coexistence scenario would point to an unconventional superconducting phase in the ground state.}

Superconductivity and magnetic order are in general mutually exclusive phenomena. Nonetheless, the coexistence of magnetism and superconductivity has been suggested first for finite momentum pairing states~\cite{FF,LO}. Coexistence of magnetism and superconductivity has been reported for 3D superconducting systems, for example, for RuSr$_2$GdCu$_2$O$_8$ \cite{Ru1212,Ru1212Pickett}, or for heavy fermion superconductors such as UGe$_2$ \cite{UGe2}. The question still remains if such a coexistence can also occur in a two-dimensional electronic system. An intriguing candidate for the coexistence of two-dimensional superconductivity and magnetic order is the interface between the two band insulators LaAlO$_3$ (LAO) and SrTiO$_3$ (STO). At their n-type interface a conducting, two dimensional, electron liquid is generated. Moreover, the LAO/STO interface was also reported to have a 2D superconducting ground state \cite{ScienceSC2007}.

For this electron liquid, magnetic ordering was suggested by Brinkman {\it et al.}, who deduced the presence of magnetic scattering centers from the temperature dependence of the interface resistance $R$ and a hysteresis of $R$ during the sweep of a magnetic field $H$~\cite{NatMaterHystMR2007}. A different magnetotransport study indicates an antiferromagnetic order~\cite{PRBMRDagan}. Also in a transport measurement, a non-uniform field-induced magnetization and strong magnetic anisotropy was found at LAO/STO interfaces~\cite{Klein}. In another recent study it was found that at both chemically treated STO bulk and LAO/STO interfaces, charges are electronically phase separated into regions containing either a quasi-two-dimensional electron gas phase, a ferromagnetic phase persisting above room temperature, or a diamagnetic/paramagnetic phase below 60 K~\cite{Ariando}. On the theoretical side, electronic structure calculations for LAO/STO interfaces have neither predicted ferromagnetism nor antiferromagnetism but yield complicated pictures for the magnetism at the interface layers \cite{PRLMillis2006, PRBPickett,Janicka,Zhong}. Specifically, for systems where the LAO surface is covered by vacuum, LDA+U calculations do not support antiferromagnetically ordered moments in the LAO/STO interface \cite{Pavlenko}. Consequently, any observed magnetism must originate from strong electronic correlations.

Coexistence of magnetism and superconductivity has not previously been reported in the LAO/STO interfaces~\cite{Dikin}. The ground state was found to be controlled by growth conditions, carrier concentration~\cite{Huijben}, and external magnetic field~\cite{Sachs}. These experimental observations based on transport properties suggest that these two phenomena do not coexist (See, e.g. Fig. 16 of Ref.~\cite{Huijben}).

To clarify whether two-dimensional superconductivity and magnetic order can coexist, we have grown LAO/STO interfaces \cite{methods}, measured their superconducting properties by transport measurements, and then applied cantilever-based torque magnetometry as an extremely sensitive and direct method to measure possible magnetic moments of the samples.

Torque magnetometry directly determines the magnetic moment $m$ of a sample by measuring the torque $\tau$ on a cantilever of the sample mounted on a cantilever is in an external magnetic field $H$.  As the torque is given by $\tau = \mathit{m} \times \mathit{B}$, the method detects the component of $m$ oriented perpendicularly to $B$. Due to its great sensitivity, this method has been applied to determine the magnetic susceptibility of very small samples, to analyze tiny magnetic signals, and, in some cases, even to accurately map Fermi surfaces~\cite{PRLanisotropyBSCCO,NatureOscillationSebastian,LiNatPhys07}.

In our setup, $\tau$ was measured with the sample glued to the tip of a 25 $\mu$m or 50 $\mu$m thick cantilever, to which $H$ was applied at a tilt angle $\phi$ with respect to the $c$-axis (perpendicular to the interface). The cantilever deflection was detected capacitively.  The magnetic moment $m$ is given by $m = \tau/(\mu_0H\sin\theta)$, where $\mu_0$ is the vacuum permeability, and $\theta$ is the angle between $m$ and $H$. (With $m$ in plane, $\theta = 90^{\mathrm{o}} - \phi$, see discussion below). Using the measured angular dependence of the zero-field capacitance of the cantilever setup, the spring constant of the cantilever was calibrated~\cite{methods}. With the spring constant, the value of $m$ can be quantitatively determined. The cantilever setup can resolve changes in $m$ of $\delta m = 10^{-13} \sim 10^{-12}$ Am$^2$ at 10 T~\cite{LiNatPhys07}.

All samples investigated were grown using nominally identical substrate preparation and pulsed laser deposition methods ~\cite{methods}. The films were patterned with Nb ohmic contacts and painted with silver paste on the back. The only intended difference between the samples is that for one reference sample (named ``0 u.c.'' sample), a shutter in front of the substrate was used to block the growth of LAO.  The resistance of the interface samples were measured using the Nb ohmic contacts. The LAO/STO interfaces were found to be superconducting below 120 mK (see discussions below).

\begin{figure}[ht]
\includegraphics[width=8.5 cm]{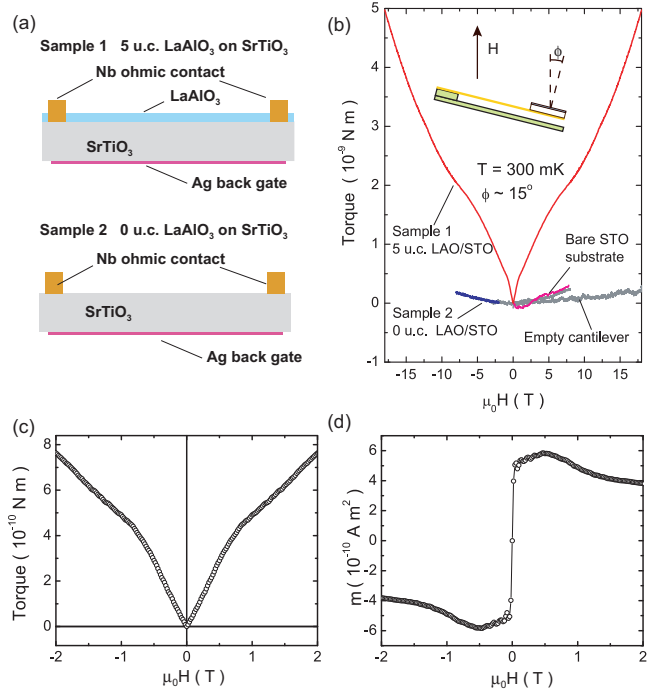}
\caption{\label{sampleTH} (color online)
Control experiments of torque signals of oxide interface systems. (Panel a) The schematics of  an  interface sample (Sample 1) and of a  0 u.c. background sample (Sample 2).The growth conditions are the same for these two groups of samples. (Panel b) The field dependence of the torque curves of various test samples (cantilever only, bare STO substrate, and the 0 u.c. sample) and the interface 5 u.c. LAO/STO sample, taken at $T$ = 300 mK and tilt angle $\phi\sim$ 15$^{\rm{\circ}}$. The inset shows a schematic of the cantilever setup. (Panel c) In Sample 1, a field dependence of the torque curve is linear and symmetric below 0.5 T. (Panel d) In Sample 1, the magnetic moment $m$ jumps to a finite value within mT next to zero field.
}
\end{figure}

An example of the $\tau-H$ dependence is shown as the red curve in Fig. \ref{sampleTH}(b) for a 5 u.c. sample. The torque signal has a pronounced reversible curve with a sharp ``cusp'' at low field.  This cusp is displayed clearly by Fig. \ref{sampleTH}(c) which zooms into this cusp. Fig. \ref{sampleTH}(d) shows the magnetic moment determined from the $\tau-H$ curve at -2 T $\leq\mu_0H\leq$ 2 T. The V-shape of the $\tau-H$ curve centered at $H$ =0 yields a nonzero, $H$-independent magnetic moment for  $\mu_0H$  up to 0.5 T. Close to $H$=0, $m$ jumps to ~ $5\times10^{-10}$ Am$^2$, corresponding to $0.3 \sim 0.4 \mu_B$ per interface unit cell (assuming that the signal is generated by the STO unit cell next to the interface, see below). The values of $m$ very close to zero field ($|\mu_0H| \leq $ 5 mT) are hard to determine, because the small $H$ causes a large relative noise in $m$. Starting at fields of order 1 Tesla, a broad dip in the $m-H$ curve suggests that an additional contribution appears in high fields. This high-field contribution was found to vary among different runs. Below we focus on the low field behavior.

\begin{figure}[ht]
\includegraphics[width=8.5 cm]{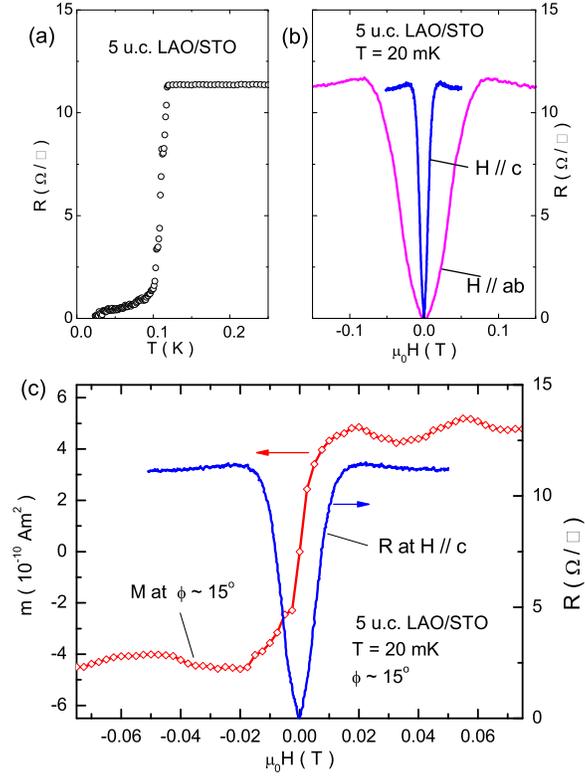}
\caption{\label{RHMHSC}  (color online)
Coexistence of superconductivity and magnetic ordering in a 5 u.c. LAO/STO interface sample. (Panel a) The dependence of the resistance $R$ on $T$ shows a superconducting transition at $T_c \sim$ 120 mK. (Panel b) Field $H$ dependence of $R$ in different field directions taken at $T$ = 20 mK. (Panel c) Field dependence of $m$ measured at $T$ = 20 mK at field at tilt angle $\phi\sim$ 15$^{\rm{\circ}}$ away from the $c$-axis. The $R-H$ curve is also plotted with $H$ parallel to the $c$-axis.
}
\end{figure}

A chief motivation for our study was to determine whether the superconductivity and magnetic order appear simultaneously or exist as separate phases in the $T-H$ phase diagram. We observe that below the superconducting $T_c$, the magnetic ordering signal and the superconducting state indeed coexist. For the sample of Fig. \ref{RHMHSC}, for example, the superconducting transition occurs at 120 mK at $H$ = 0, with a resistance foot extending to 25 mK. The $R - H$ curves measured at 20 mK with $H$ parallel and perpendicular to the interface plane are plotted in Fig. \ref{RHMHSC}(b). While the interface is superconducting, the $m-H$ curve at 20 mK displays the same jump at small fields (Fig. \ref{RHMHSC}(c)) as that observed at higher temperatures (Fig.~\ref{sampleTH}(d)). Notably, a finite magnetic moment is recorded at $\mu_0H\sim$  5 mT, while the sample resistance $R$ does not reach the normal state values until $\mu_0H\sim$ 20 mT. The magnetic ordering signal and the superconducting state are therefore found to coexist.

\begin{figure}[ht]
\includegraphics[width=8.5 cm]{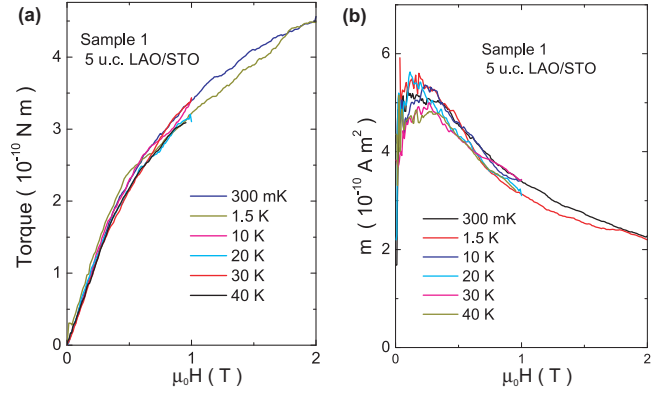}
\caption{\label{highT} (color online)
Torque$-H$ and $m-H$ curves measured at elevated temperatures (tilt angle $\phi\sim$ 49$^{\rm{\circ}})$.
}
\end{figure}

The magnetic ordering signal is robust at elevated temperatures. For the 5 u.c. LAO/STO sample, the magnetic moment does not show a significant temperature dependence even up to 40\,K(Fig. \ref{highT}), the highest $T$ at which his sample was investigated. In another 5 u.c. LAO/STO sample, $m$ was found to be nonzero up to 200 K~\cite{methods}. Such a $T$-dependence is consistent with previous results~\cite{Ariando}, reporting the existence of ordering state at room temperature. The samples are therefore found to have a high magnetic ordering temperature, indicating a strong magnetic exchange coupling.

The magnetic field dependence of $m$ can be described by the Langevin-function characteristic for superparamagnetism, where spins are aligned in small size domains to behave as a large classical magnetic moment~\cite{bookMorrish}. However, superparamagnetic samples usually show a strong temperature dependence in the low field $m-H$ curves, a feature missing in the $m-H$ curves in Fig. \ref{highT}. The reason of this missing is probably due to the noise of $m$ at fields close to zero. Since up to at least 40 K the measured moment $m$ saturates at about 30 mT , the lower bound of the collective classical moment is around $\sim 10^3\mu_B$.  On the other hand, the $m-H$ curves are also consistent with a very soft ferromagnet whose hysteresis loop is hidden by the $m$ noise at $|\mu_0H| <$ 5 mT,  Although these two possibilities cannot be distinguished by our data, all of them suggest a strong ferromagnetic-like magnetic coupling within magnetic domains.

To determine the orientation of the magnetic moment, a series of torque measurements was performed in which the angle of the sample relative to $H$ was varied (see inset in Fig. \ref{sampleTH}). Because $\tau = \mathbf{m}\times\mathbf{B} = \mu_0mH_{\bot}$, where $H_{\bot}$ is the component of $H$ perpendicular to the magnetic moment $m$, the orientation of the moments can be discerned by tracking the angular dependence of the torque signal. In highly anisotropic system like the interface, $m$ is determined by $H_{\parallel}$, the field component parallel to $m$. Thus if  $H_{\parallel}$  is large enough to saturate $m$, any change of $\tau$ is generated by the change of $H_{\bot}$ alone, and $\tau$ will increase with the sine of the angle between $H$ and $m$. On the other hand, once $H_{\parallel}$ is insufficient to saturate $m$, the torque will stop following the sine behavior.

\begin{figure}[ht]
\includegraphics[width=8.5 cm]{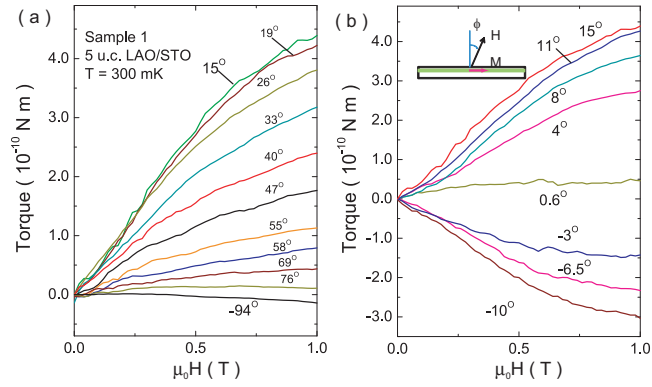}
\caption{\label{Angular} (color online)
Torque curves measured at different tilt angles in the 5 u.c. LAO/STO sample.
}
\end{figure}

The angle-dependent torque measurements show that the low-field saturation magnetic moment stays in the plane of the interface. We carried out low-field torque measurements at 300~mK at 30 different tilt angles. Fig. \ref{Angular} shows the $\tau-H$ curves at several selected angles $\phi$. As shown in Fig. \ref{Angular}(a), as $\phi$ changes from 15$^{\rm{\circ}}$ to 94$^{\rm{\circ}}$, $\tau$ decreases monotonically and slowly approaches zero at $\phi =$ 90$^{\rm{\circ}}$, where the $H$ direction is almost parallel to the direction of $m$. On the other hand, as $\phi$ varies between $+15^{\rm{\circ}}$ and $-10^{\rm{\circ}}$, the applied field $H$ is almost perpendicular to the direction of $m$. The in-plane projection of the field $H_{ab}$ decreases and eventually changes to the opposite direction. The in-plane $m$ drops to zero once $H_{\parallel}$ is close to zero. As a result, the $\tau$ ($H$) curves swing from a positive saturation at $\phi\sim 15^{\rm{\circ}}$ to a negative saturation at $\phi\sim -10^{\rm{\circ}}$.

To explore whether the recorded torque signals originate from the LAO/STO interface or from other sources, several control experiments were performed by using reference samples, including an empty cantilever.  Sizable torque signals were only observed from samples containing LAO/STO interfaces, whose torque exceeds that of all other background samples by two orders of magnitude. We also note that for all background contributions, such as possible magnetic signal arising from the standard bulk STO, standard bulk LAO, and the Ag back gate, $m$ will be proportional to $H$, as these materials are paramagnetic or diamagnetic. To explore whether the moment is generated by defects of the LAO film, a 5 u.c. thick LAO film was grown on a LAO substrate. The measured torque signal is again two orders of magnitude smaller~\cite{methods} than that of the 5 u.c. LAO/STO sample.  While the Nb ohmic contacts are superconducting below 9 K, they are not the source of the torque signal, as the torque is found far above the upper critical field of Nb (0.4 T at 0~K).  Further, the torque magnetometry is only sensitive to the transverse magnetic response, while all background magnetic moments will be oriented closely parallel to $H$. We therefore conclude that the observed large torque indeed arises from the presence of the LAO/STO interface.

Our data show that 2D superconductivity and magnetic order coexist at n-type LAO/STO interfaces. The results leave the question open on whether the same electrons are generating the superconducting and the magnetic order. The fact that the magnetic response is superparamagnetic-like and the superconductivity causes a vanishing resistance is consistent with scenarios in which ferromagnetic order is present in small puddles formed as a separate phase in a percolating, superconducting matrix. The measured results can also be accounted for by scenarios of vertical domains, in which laterally inhomogeneous magnetic and superconducting electron layers are generated in the STO at different depths away from the interface. In both cases, the superconducting phase is in close contact to the ferromagnetic phase, so that a large volume of the superconducting phase will likely be affected by the ferromagnetism. Furthermore, the data are also consistent with the idea that the same electron system forms an inhomogeneous, magnetically ordered, superconducting electron liquid. This notion is in accord with a proposal that the oxygen vacancies in the interfacial TiO$_2$ layers stabilize ferromagnetic-type order in the Ti ions close to the interface, as supported by DFT-calculations~\cite{Natalja}.

In conclusion, by magnetic torque magnetometry we have directly shown the presence of magnetic order in the two-dimensional electron liquid of LAO/STO interfaces. The magnetic order is characterized by a superparamagnetic-like behavior, with saturation magnetic moments of $\sim$ 0.3 $\mu_B$ per interface unit cell oriented in-plane, persisting beyond 200 K. Below 120 mK, the ferromagnetic-like magnetic order coexists with the 2D-superconductivity.

{\it{Materials and Methods}}

The LaAlO$_3$/SrTiO$_3$ heterostructures were grown at Augsburg University using pulse-laser deposition with in-situ monitoring of the LaAlO$_3$ layer thickness by reflection high-energy electron diffraction. The single crystalline SrTiO$_3$ substrates were TiO$_2$ terminated. Their lateral size is  $5 \times 5$ mm$^2$ and their thickness is 1 mm. The LaAlO$_3$ layers were grown at an oxygen pressure of $8\times10^{-5}$ mbar at 780 $^\mathrm{o}$C to a thickness of 5 u.c. The sputtered ohmic Nb contacts filled holes patterned by etching with an Ar ion-beam. The reference (0 u.c. ) samples were grown in the same conditions (oxygen pressure of $8\times10^{-5}$ mbar at 780 $^\mathrm{o}$C).

The magnetization measurements were preformed with a home-built cantilever-based torque magnetometry apparatus at MIT. Cantilevers are made from thin gold or brass foils. We deposit gold film on a sapphire and put it under the cantilever. The torque is tracked by measuring the capacitance between the cantilever and the gold film, using a GR1615 capacitance bridge or an AH2700A capacitance bridge. To calibrate the spring constant of the cantilever, we rotate the cantilever setup under zero magnetic field to measure the capacitance change caused by the weight of the sample wafer.

\newpage
{\it{Acknowledgment}}

The authors gratefully acknowledge helpful discussions with T. Kopp, P. A. Lee, and G. A. Sawatzky. This work was supported by ARO-54173PH, by the National Science Foundation through the NSEC program, by the DFG (TRR 80), the EC (OxIDes), and by the Nanoscale Research Initiative.  L. Li would like to thank the MIT Pappalardo Fellowships in Physics for their support. The high-field experiments were performed at the National High Magnetic Field Laboratory, which is supported by NSF Cooperative Agreement No. DMR-084173, by the State of Florida, and by the DOE.

\newpage

%

\end{document}